\documentclass[aps,twocolumn,prl]{revtex4-1}

\usepackage{graphicx}% Include figure files
\usepackage{dcolumn}% Align table columns on decimal point
\usepackage{bm}% bold math
\usepackage{xcolor}
\usepackage{amssymb}   % for math
\usepackage{amsmath}
\usepackage{ulem}

\begin{document}
\title{
Unusual thermal Hall effect in a Kitaev spin liquid candidate $\alpha$-RuCl$_3$
}

\author{Y. Kasahara$^1$}
\altaffiliation{These authors contributed equally to this work.}
\author{K. Sugii$^{2,\ast}$}
\author{T. Ohnishi$^1$}
\author{M. Shimozawa$^2$}
\author{M. Yamashita$^2$}
\author{N.~Kurita$^3$}
\author{H. Tanaka$^3$}
\author{J. Nasu$^3$}
\author{Y. Motome$^4$}
\author{T. Shibauchi$^5$}
\author{Y. Matsuda$^1$}

\affiliation{$^1$Department of Physics, Kyoto University, Kyoto 606-8502, Japan}
\affiliation{$^2$Institute for Solid State Physics, University of Tokyo, Kashiwa 277-8581, Japan}
\affiliation{$^3$Department of Physics, Tokyo Institute of Technology, Meguro, Tokyo 152-8551, Japan}
\affiliation{$^4$Department of Applied Physics, University of Tokyo, Bunkyo, Tokyo 113-8656, Japan}
\affiliation{$^5$Department of Advanced Materials Science, University of Tokyo, Chiba 277-8561, Japan}

\begin{abstract}
The Kitaev quantum spin liquid displays the fractionalization of quantum spins into Majorana fermions. The emergent Majorana edge current is predicted to manifest itself in the form of a finite thermal Hall effect, a feature commonly discussed in topological superconductors. Here we report on thermal Hall conductivity $\kappa_{xy}$ measurements in $\alpha$-RuCl$_3$, a candidate Kitaev magnet with the two-dimensional honeycomb lattice. In a spin-liquid (Kitaev paramagnetic) state below the temperature characterized by the Kitaev interaction $J_K/k_B \sim 80$\,K, positive $\kappa_{xy}$ develops gradually upon cooling, demonstrating the presence of highly unusual itinerant excitations. Although the zero-temperature property is masked by the magnetic ordering at $T_N=7$\,K, the sign, magnitude, and $T$-dependence of $\kappa_{xy}/T$ at intermediate temperatures follows the predicted trend of the itinerant Majorana excitations. 
\end{abstract}

\maketitle
Quantum spin liquids (QSLs) are novel states of matter lacking magnetic order while possessing some special patterns of quantum mechanical entanglement \cite{Balents10}. Among the long-standing experimental challenges associated with the identification of these exotic states, is the detection of fractionalized  excitations, which are signatures of  topological order inherent to the QSL. Recently, the Kitaev spin model of insulating magnets on a two-dimensional (2D) honeycomb lattice has attracted interest, as it hosts a QSL where spins are fractionalized into two types of Majorana fermions; one is localized and composes localized $Z_2$ fluxes with finite gap of $\Delta_G\sim 0.06 J_K$, and the other is itinerant (mobile) and gapless at zero field [Fig.\,1(a)] \cite{Kitaev06,Trebst17}.    

A spin-orbital-assisted Mott insulator $\alpha$-RuCl$_3$, in which local  $j_\mathrm{eff}=1/2$ spins  are aligned in 2D honeycomb layers, has been recently considered to be a Kitaev candidate material [Fig.\,1(b)]. Although $\alpha$-RuCl$_3$ exhibits an antiferromagnetic (AFM) ordering with zigzag spin structure at $T_N \approx7$\,K  \cite{Johnson15} due to non-Kitaev interactions, such as Heisenberg exchange and off-diagonal interactions [Fig.\,1(c)], $J_K$ is predominant among the magnetic interactions \cite{Plumb14,Kim15,Jackeli09,Banerjee16,Sandilands15,Nasu16}.   The signature of the spin liquid, in which spins are not ordered but strongly entangled due to Kitaev interaction, is expected to appear in the temperature range between $T_N$ and $J_K/k_B$. The unusual broad magnetic excitations reported by Raman and neutron inelastic scattering measurements have been suggested to reflect the proximity to the Kitaev model \cite{Banerjee16,Sandilands15,Nasu16,Yoshitake16}.  However, the interpretation of these excitations in terms of itinerant Majorana fermions is controversial \cite{Trebst17,Winter17}. Thus, the further investigation of excitations is strongly desired. 

Thermal transport experiments constitute sensitive probes to study low-energy itinerant quasiparticle excitations, as the localized degrees of freedom, such as nuclear spins and defects, do not carry heat. In conventional insulating magnets, the longitudinal thermal conductivity is observed due to bosonic magnon excitations in the ordered state, but it is strongly suppressed in the conventional paramagnetic state.   On the other hand, in the disordered QSL states realized in 1D chain, and 2D triangular and 3D pyrochlore systems,  large thermal conductivity is observed, which has been attributed to  fractionalized exotic quasiparticles,  such as spinons and monopoles \cite{Sologubenko01,Yamashita10,Tokiwa16}.    Recently, the thermal Hall effect, thermal analogue of the electronic Hall effect, has aroused great attention as a unique probe of non-trivial topological excitations \cite{Katsura10,Onose10,Ideue12,Hirschberger15,Watanabe16, Han16}. In particular, it has been predicted that in the spin-liquid state of Kitaev magnets under an external field thermal Hall effect can be generated by Majorana fermion excitations characterized by the non-zero topological Chern number. Therefore, thermal Hall effect is a powerful probe in the search for Majorana fermions as commonly discussed in quantum Hall systems and chiral $p$-wave superconductors \cite{Read00,Nomura12,Sumiyoshi13}. 

Here we report the observation of a large $\kappa_{xy}$ with positive sign in the spin-liquid (Kitaev paramagnetic) state of $\alpha$-RuCl$_3$ at $T_N<T<J_K/k_\mathrm{B}$. We find that the observed thermal Hall response is essentially different from the other insulating magnets where finite $\kappa_{xy}$ has been reported, demonstrating the presence of highly unusual itinerant excitations inherent to the spin-liquid state of $\alpha$-RuCl$_3$.  

 \begin{figure}[t]
 	\begin{center}
 		\includegraphics[width=1.0\linewidth]{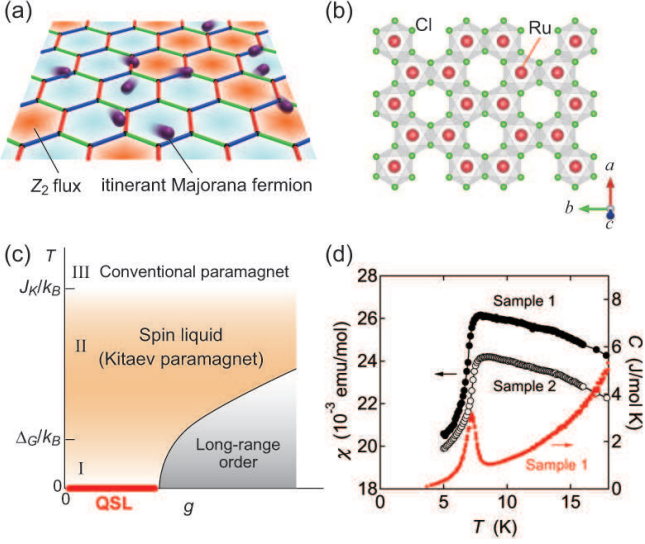}
 		\caption{
 			(a) Schematic of Kitaev model on honeycomb lattice. Blue, green, and red bonds represent Ising-like bond-directional interactions between $x$-, $y$-, and $z$-components of the 1/2 spins at the apexes of the hexagons, respectively. The quantum spins are fractionalized into itinerant Majorana fermions (purple spheres) and $Z_2$ fluxes with $W_p=\pm 1$ (orange hexagons are excited fluxes for $W_p=-1$).  
 			(b) Crystal structure of $\alpha$-RuCl$_3$ in the $ab$ plane.  Edge-sharing RuCl$_6$ octahedra forms 2D honeycomb lattice of Ru ions [Ru$^{3+}(4d^5)$, $j_\mathrm{eff}=1/2$].  
 			(c) Schematic phase diagram for the 2D Kitaev model as a function of $g$, where $g$ is the ratio of the Kitaev interaction $J_K$ and non-Kitaev interactions, such as Heisenberg and off-diagonal terms (in the pure Kitaev model, $g=0$).  At low $g$, the ground state is a QSL.   $\alpha$-RuCl$_3$ corresponds to finite $g$ with magnetic ordering.   There are two characteristic temperatures, $J_K/k_\mathrm{B}$ and $\Delta_G/k_B \sim 0.06 J_K/k_B$.  At $T>J_K/k_B$ (regime-III), the system is in a conventional paramagnetic state.  At $T<J_K/k_B$, the spin-liquid (Kitaev paramagnetic) state appears. Above $\Delta_G/k_B$ (regime-II), the itinerant Majorana fermions are scattered by the thermally excited $Z_2$ fluxes. Below $\Delta_G/k_B$ (regime-I),  the numbers of the excited $Z_2$ fluxes are strongly reduced. 
			(d) Temperature dependence of magnetic susceptibility $\chi$ (left axis) measured in an in-plane magnetic field of $H=1000$\,Oe for two different $\alpha$-RuCl$_3$ single crystals and specific heat $C$ in zero field (right axis) for sample 1. 
 		}
 	\end{center}
 \end{figure}
 
 \begin{figure}[t]
 	\begin{center}
 		\includegraphics[width=0.75\linewidth]{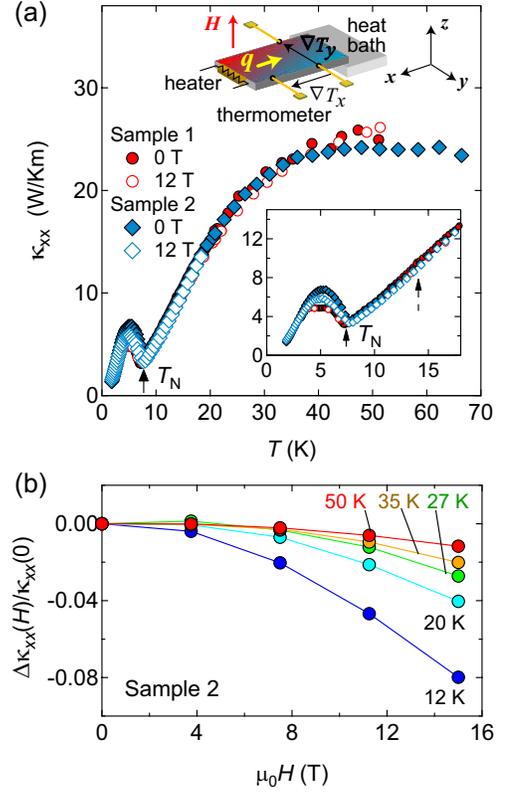}
 		\caption{
 			(a) Temperature dependence of  $\kappa_{xx}$ at 0\,T and 12\,T for two different samples.   Lower inset shows expanded view of $\kappa_{xx}(T)$ at low temperatures. Solid arrows indicate the AFM transition temperature $T_N\sim7$\,K.  No anomaly is observed at 14\,K at which the transition associated with the stacking fault occurs (dashed arrow).  Upper inset illustrates a schematic of the measurement setup for the  in-plane thermal conductivity and thermal Hall conductivity. 
 			(b) Field dependence of magneto-thermal conductivity $\Delta\kappa_{xx}(H)/\kappa_{xx}(0)[\equiv \frac{\kappa(H)-\kappa(0)}{\kappa_{xx}(0)}]$ in the spin-liquid state.  
 		}
 	\end{center}
 \end{figure}

High-quality single crystals of $\alpha$-RuCl$_3$ were grown by a vertical Bridgman method \cite{Kubota15}. In $\alpha$-RuCl$_3$, another magnetic transition due to the stacking faults often occurs at around 14\,K, where magnetic susceptibility and specific heat exhibit a clear kink and a sharp peak, respectively \cite{Cao16}. For the thermal transport measurements, we  selected single crystals in which specific heat shows sharp anomaly at 7\,K but shows no anomaly at 14\,K [Fig.\,1(d)].  As the bending of the crystal may induce the stacking faults \cite{Cao16}, we remeasured the specific heat after the thermal transport measurements and confirmed the absence of magnetic transition at 14\,K, implying the minimal effect of stacking fault during the measurements. The sample size (distance between contacts for longitudinal thermal gradient $\nabla_xT$) is $2\times0.5\times0.02$\,mm$^3$ (1.2\,mm) and $2\times0.7\times0.03$\,mm$^3$ (1.1\,mm) for Sample 1 and 2, respectively. In-plane thermal conductivity and thermal Hall conductivity were measured simultaneously on the same crystal by the standard steady state method in high vacuum, using the experimental setup illustrated in the inset of Fig.\,2(a).  The samples are suspended between heat bath and a chip resistance heater.  Heat current $\bm{q}$ and magnetic field $\bm{H}$ were applied along the $ab$ plane ($\bm{q}$$\parallel$$\bm{x}$) and $c$ axis ($\bm{H}$$\parallel$$c$$\parallel$$\bm{z}$), respectively. The temperature gradients $-\nabla_xT$$\parallel$$\bm{x}$ and $-\nabla_yT$$\parallel$$\bm{y}$ were measured by three Cernox thermometers carefully calibrated under magnetic fields. For the measurements of the thermal Hall effect, we removed the longitudinal response due to misalignment of the contacts by anti-symmetrizing the measured $\nabla_yT$ as $\nabla_yT^\mathrm{asym}(H)=[\nabla_yT(H)-\nabla_yT(-H)]/2$ at each temperature. $\kappa_{xx}$ and $\kappa_{xy}$ were obtained from longitudinal thermal resistivity $w_{xx}=\nabla_xT/q$ and thermal Hall resistivity $w_{xy}=\nabla_yT^\mathrm{asym}/q$ as $\kappa_{xx}=w_{xx}/(w_{xx}^2+w_{xy}^2)$ and $\kappa_{xy}=w_{xy}/(w_{xx}^2+w_{xy}^2)$. We performed measurements with different applied heat current and confirmed linear responses in $\nabla_xT$ and $\nabla_yT^\mathrm{asym}$. To avoid background Hall signal, a LiF heat bath and non-metallic grease were used \cite{Watanabe16}. The thermal transport measurements were performed on two different single crystals in the same batch, samples 1 and 2, at Kyoto Univ. and at Univ. of Tokyo, respectively, by using similar setups with different calibrated thermometers (Cernox CX1070 and CX1050 bare chips were used in the setups at Kyoto and at Tokyo, repspectively). As shown in Figs.\,2(a) and 4(a), both $\kappa_{xx}(T)$ and $\kappa_{xy}(T)$ of sample 1 well coincide with those of sample 2. 

Figure\,2(a) depicts the in-plane thermal conductivity $\kappa_{xx}(T)$ of $\alpha$-RuCl$_3$ single crystals in zero and finite magnetic fields ($H$) perpendicular to the 2D planes.   The overall $T$-dependence is similar to the previous reports \cite{Hentrich16,Hirobe16,Leahy16}.  At $T_N$, $\kappa_{xx}$ exhibits a sharp kink anomaly.  Although the magnetic susceptibility shows a tiny anomaly at 14\,K due to the transition associated with stacking faults in the crystals [Fig.\,1(d)] \cite{Kubota15,Sears}, $\kappa_{xx}$ and specific heat exhibit no corresponding anomaly [Figs.\,1(d) and 2(a)], indicating that this effect is negligible in the present analysis. As $\alpha$-RuCl$_3$ is a good insulator, in which electrical resistivity is hard to measure even at room temperature, mobile charge carriers are absent. In insulating spin systems, impurities usually act as scattering centers that reduce $\kappa_{xx}$ regardless of magnetic and non-magnetic ones. Therefore, the fact that the magnitude of $\kappa_{xx}$ in the present crystals is 2-3 times larger than the previously reported values suggest the high quality of the present crystals with smaller impurity/disorder levels and longer mean free path of heat carriers.

 \begin{figure}[t]
 	\begin{center}
 		\includegraphics[width=1.0\linewidth]{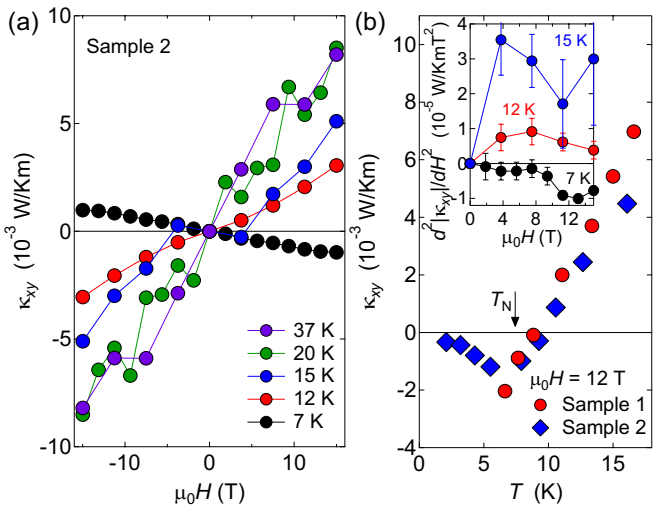}
 		\caption{
 			(a) Field dependence of $\kappa_{xy}$ for sample 2. 
			(b) Temperature dependence of $\kappa_{xy}$  near $T_N$. 
			Inset shows the field dependence of $d^2|\kappa_{xy}(H)|/dH^2$ below and above $T_N$.
 		}
 	\end{center}
 \end{figure}

As shown in Fig.\,2(b), $\kappa_{xx}(H)$ decreases with $H$.  The suppression of $\kappa_{xx}(H)$ in perpendicular field  is much weaker than that in parallel field, which is due to the strong magnetic anisotropy of $\alpha$-RuCl$_3$ \cite{Majumder15} (Although not shown here, our results in parallel fields are essentially the same as those reported in \cite{Hentrich16,Leahy16}). In the present system, the heat is carried by spin excitations and phonons;  $\kappa_{xx}=\kappa_{xx}^\mathrm{sp}+\kappa_{xx}^\mathrm{ph}$.  As the magnetic field enhances the phonon mean free path due to the alignment of spins,  suppressing the spin-phonon scattering rate,  $\kappa_{xx}^\mathrm{ph}$ is expected to increase with $H$.   Thus the observed reduction of $\kappa_{xx}(H)$ with $H$ is caused by  $\kappa_{xx}^\mathrm{sp}$.  However, lacking detailed information on spin-phonon scattering, which is suggested  to be present in this material \cite{Hentrich16}, the quantitative contribution of spin excitations is difficult to be extracted from $\kappa_{xx}(H)$. Thermal Hall effect is, therefore, a more suitable probe to unveil the nature of spin excitations. 

\begin{figure}[t]
 	\begin{center}
 		\includegraphics[width=1.0\linewidth]{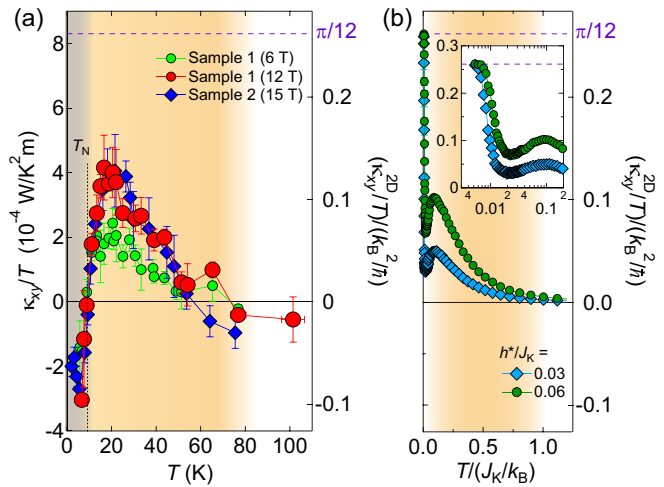}
 		\caption{
 			(a) Temperature dependence of $\kappa_{xy}/T$ of $\alpha$-RuCl$_3$ in several magnetic fields in the spin-liquid and magnetically ordered states.   The right scale represents  $(\kappa_{xy}^\mathrm{2D}/T)/(k_\mathrm{B}^2/\hbar)$, where  $\kappa_{xy}^\mathrm{2D}=\kappa_{xy}d$ ($d=5.72$\,\AA\ is the interlayer distance \cite{Johnson15}).  Violet dashed line represents the quantized thermal Hall effect; $(\kappa_{xy}^\mathrm{2D}/T)/(k_\mathrm{B}^2/\hbar)=\pi/12$.
 			(b) Numerical results of $(\kappa_{xy}^\mathrm{2D}/T)/(k_\mathrm{B}^2/\hbar)$ plotted as a function of $T/(J_K/k_B)$ for the pure Kitaev model in the presence of the effective magnetic field $h^\ast$ proportional to $H^3$, which is introduced by treating $H$ as a perturbation \cite{Nasu17}. Inset shows the same plot on a semi-logarithmic scale of $T/(J_K/k_B)$.  Regimes I and II represent the $T$-regime below and above $\Delta_G/k_B$ in the spin-liquid state, respectively,  and regime III represents the conventional paramagnetic state.  
			}
 	\end{center}
 \end{figure}

As shown in Fig.\,3(a), $\kappa_{xy}$, obtained by anti-symmetrizing thermal response with respect to the field direction, is clearly resolved, establishing a finite thermal Hall effect.  Since the magnitude of $\kappa_{xy}$ is three orders of magnitude smaller than $\kappa_{xx}$ even at 12\,T, special care was taken to detect the intrinsic thermal Hall signal \cite{Watanabe16}.  Figure\,3(b) and its inset depict the $T$-dependence of $\kappa_{xy}$ and $H$-dependence of $d^2|\kappa_{xy}|/dH^2$, respectively, below and above $T_N$.  In the spin-liquid state above $\sim T_N$, $\kappa_{xy}$ is positive and displays a superlinear field dependence ($\kappa_{xx}\propto H^\alpha$ with $\alpha>1$ and $d^2|\kappa_{xy}|/dH^2>0$), while it becomes negative and exhibits a sublinear field dependence with saturating behavior at high $H$ ($\kappa_{xx}\propto H^\alpha$ with $\alpha<1$ and $d^2|\kappa_{xy}|/dH^2<0$) upon entering the ordered state.  The sign changes in both $\kappa_{xy}$ and $d^2|\kappa_{xy}|/dH^2$ when crossing around $T_N$ suggest that the thermal Hall effect in the ordered and spin-liquid states is different in origin. Figure\,4(a) displays $\kappa_{xy}/T$ in a wide temperature range.  Above $T_N$, $\kappa_{xy}/T$ increases steeply with $T$ and decreases gradually after reaching a maximum at around 20\,K. No discernible  Hall signal  is observed within our experimental resolution in the conventional paramagnetic state above $\sim80$\,K. 

In insulators with no charge carrier, there are several possible origins for finite $\kappa_{xy}$, including phonons, magnons, and exotic spin excitations.  We point out that both phonons and magnons are unlikely because of the following reasons.  According to previous reports, the phonon thermal Hall conductivity  $\kappa_{xy}^{ph}/T$  is two orders of magnitude smaller than the observed $\kappa_{xy}/T$.  Moreover, $\kappa_{xy}^{ph}/T$ shows essentially different $T$-dependence \cite{Strohm05,Sugii17}. The magnon thermal Hall effect, which is induced by the Berry curvature effect, appears only in the ferromagnetically ordered state \cite{Onose10,Ideue12,Hirschberger15}. In spin liquid states, it has been predicted that the spin chirality arising from the Dzyaloshinsky-Moriya (DM) interaction gives rise to finite $\kappa_{xy}$ \cite{Han16,Owerre,Kovalev}. Experimentally, finite $\kappa_{xy}$ in a spin-liquid state has been reported only in kagom\'{e} volborthite Cu$_3$V$_2$O$_7$(OH)$_2\cdot$2H$_2$O with a large DM interaction.  However,  DM interaction in $\alpha$-RuCl$_3$, which is allowed by symmetry for the second neighbor bonds in the honeycomb lattice, is very small, $D/k_B\sim 5$\,K $\approx 0.06J_K/k_B$ \cite{Winter16}.  Therefore, the fact that  the prominent positive $\kappa_{xy}/T$ is observable up to $\sim J_K/k_B$ in $\alpha$-RuCl$_3$ suggests that the spin chirality is unlikely as an origin of finite $\kappa_{xy}/T$.  We note that in volborthite the Hall sign is negative and the amplitude of $\kappa_{xy}/T$ is one order of magnitude smaller than that in $\alpha$-RuCl$_3$ \cite{Watanabe16}. The thermal Hall effect is theoretically predicted in a spin-liquid state with spinon Fermi surface through the coupling with a gauge field \cite{Katsura10}.  However, it has been reported that in $\alpha$-RuCl$_3$ the low-lying gapless fermionic excitations are absent, indicating no spinon Fermi surface \cite{Sears17,Yu17}. These results lead us to conclude that the observed thermal Hall effect above $T_N$ reflects unusual quasiparticle excitations inherent to the spin-liquid state of $\alpha$-RuCl$_3$.
 
In the Kitaev magnets the thermal Hall response is dominated by the itinerant Majorana fermions. The sign of $\kappa_{xy}$ is determined by the topological Chern number and is predicted to be positive \cite{Kitaev06}, which is consistent with the present results.  Moreover, the non-linear $H$-dependence with upward curvature of $\kappa_{xy}(H)$ is predicted \cite{Nasu17}. This appears to be consistent with the observed $\kappa_{xy}(H)$ at 12\,K above $T_N$, although non-linear $H$-dependence is less clear at $T>15$\,K.   In the zero-temperature limit,  the $T$-linear coefficient of the thermal Hall conductivity per 2D layer $\kappa_{xy}^\mathrm{2D}/T$  is quantized as $\kappa_{xy}^\mathrm{2D}/T=(\pi/12)(k_\mathrm{B}^2/\hbar)$ \cite{Kitaev06,Nasu17}. The right axis of Fig.\,4(a) plots $(\kappa_{xy}^\mathrm{2D}/T)/(k_B^2/\hbar)$, where  $\kappa_{xy}^\mathrm{2D}=\kappa_{xy}d$ ($d$ is the interlayer distance).   The result is highlighted by the fact that $(\kappa_{xy}^\mathrm{2D}/T)/(k_B^2/\hbar)$ at 20\,K reaches about a half of $\pi/12$ at $T\rightarrow 0$.  Considering the occurrence of the magnetic ordering and the presence of non-Kitaev terms, this observation is quite telling of the nature of the spin-liquid phase of this material. 

Figure\,4(b) displays the numerical results of $\kappa_{xy}(T)/T$ for the 2D pure Kitaev model calculated by the quantum Monte Carlo method \cite{Nasu17}. In $\alpha$-RuCl$_3$, where each layers are weakly coupled by van der Waals interaction, the exchange coupling between the layers is very weak (two orders of magnitude smaller than the in-plane exchange interactions) \cite{Kim16}, which demonstrates the 2D nature of magnetic properties. In the numerical results, there are three distinct temperature regimes.  In the high-temperature conventional paramagnetic state (regime-III), $\kappa_{xy}/T$ vanishes.  In the spin-liquid state below $\sim J_K/k_B$ (regime-II),  $\kappa_{xy}(T)/T$ is generated by itinerant Majorana fermions, which are scattered by the thermally excited $Z_2$ fluxes.  A broad peak of $\kappa_{xy}/T$ at $T\sim 0.1 J_K/k_B$ appears as a result of the interaction between the two excitations; itinerant Majorana fermions and $Z_2$ fluxes [Fig.\,1(a)].  At low temperatures at $T<\Delta_G/k_B$ (regime-I),  the rapid reduction of the number of the excited $Z_2$ fluxes leads to a steep increase of $\kappa_{xy}/T$, approaching the quantized value at $T\rightarrow0$.  The numerical calculations do not take into account the non-Kitaev terms, and thus the occurrence of magnetic ordering. Apart from the low-temperature range where the effects of magnetic order take place, the experimental curve of $\kappa_{xy}(T)/T$ follows the trend of the calculated results. AFM fluctuations above $T_N$ would suppress the Kitaev excitations, which may be partly responsible for the reduction of $\kappa_{xy}(T)/T$ near $T_N$. However, whether such an effect is significant up to much higher temperature than $T_N$ is not clear.   Although it should be scrutinized by other probes, the fact that the thermal conductivity, magnetic susceptibility and specific heat show no critical features associated with the transition except for very near $T_N$, suggests that the experimentally observed peak of $\kappa_{xy}/T$ at much higher temperature $\sim$20\,K ($\sim0.25J_K/k_B$) likely has the same origin as the peak feature found in the calculation.  Even if the AFM  fluctuations originating from the ordered state below $T_N$ influence the thermal Hall signal, we can safely conclude that $\kappa_{xy}/T$ in the high temperature regime above $\sim$20\,K, where $\kappa_{xy}/T$ increases with decreasing $T$, is most likely to be attributed to unusual spin excitations inherent to the Kitaev spin liquid. 

Although the zero-temperature property is masked by the magnetic ordering, the magnitude of thermal Hall signal as well as its sign and its $T$-dependence above $T_N$, including a peak structure are well reproduced by the Monte Carlo calculation. There is no calculation of thermal Hall effect in which non-Kitaev interactions are taken into account in the spin liquid state at finite temperature. Thus, at the present stage, we have shown that the pure Kitaev model provides a plausible explanation for the observed thermal Hall signal. Our results appear to provide a signature of the Majorana fermion excitations in the spin-liquid (Kitaev paramagnetic) state, where itinerant topological Majorana fermions propagate in the background of thermally-excited $Z_2$ fluxes.

In summary, we have investigated the low-energy itinerant quasiparticle excitations in the Kitaev spin liquid candidate $\alpha$-RuCl$_3$ by thermal Hall effect. Our measurements reveal a large thermal Hall signal as well as the anomalous $T$-dependence in the spin-liquid state of $\alpha$-RuCl$_3$, which is markedly different from the other insulating magnets. Although the anomalous behavior is masked by the magnetic ordering below $T_N$, the sign, magnitude, and $T$-dependence of the observed $\kappa_{xy}/T$ above $T_N$ are well reproduced by the calculated $\kappa_{xy}/T$ for the 2D pure Kitaev model, suggesting that the observed thermal Hall response arises from quasiparticle excitations inherent to the Kitaev spin liquid. 

%   \section*{Acknowledgements}
 We thank S. Fujimoto, H. Katsura, E.-G. Moon, H. Lee, K. Nomura, Y. Tokiwa, K. Totsuka, and M. Udagawa, for useful discussions. This work was supported by Grants-in-Aid for Scientific Research (KAKENHI) (No.\,25220710, 15H02014, 15H02106, and 15H05457) and Grants-in-Aid for Scientific Research on innovative areas ``Topological Materials Science" (No.\,JP15H05852) from Japan Society for the Promotion of Science (JSPS), and by Yamada
Science Foundation and Toray Science Foundation.

\end{document}